# Fulfillment Request Management (The approach)

Stylianos Gkoutzioupas

Department of Business Administration, School of Buisiness, Univercity of the Aegean, 8 Micahlon Str. Chios Greece

*ABSTRACT*

*In this paper we introduce the term FRM (Fulfillment Request Management). According to the FRM in a BSS / OSS environment we can use a unified approach to implement a SOA in order to integrate BSS with OSS and handle 1. Orders 2. Events 3. Processes. So in a way that systems like ESB, Order Management, and Business Process Management can be implemented under a unified architecture and a unified implementation. We assume that all the above mentioned are 'requests' and according to the system we want to implement, the request can be an event, an order, a process etc. So instead of having N systems we have 1 system that covers all the above (ESB, Order Management, BPM etc) With the FRM we can have certain advantages such as: 1. adaptation 2. Interoperability. 3. Re-usability 4. Fast implementation 5. Easy reporting. In this paper we present a set of the main principles in order to build an FRM System.*

## 1. INTRODUCTION

BSS/OSS

In telecom organizations the heart of Computer and Information Technology systems are BSS/OSS. BSS stands for "Business Support Systems" and OSS for "Operation Support Systems". Business Support Systems (BSS) refer to "Business Layer" inside the organization and are responsible for customer entry, order entry, billing and invoicing of the customer, payment collection etc. Operations Support Systems (OSS) refer to "Network Layer" inside the organization and are responsible for service provisioning, network inventory, network element configuration, fault management etc.

Figure 1.

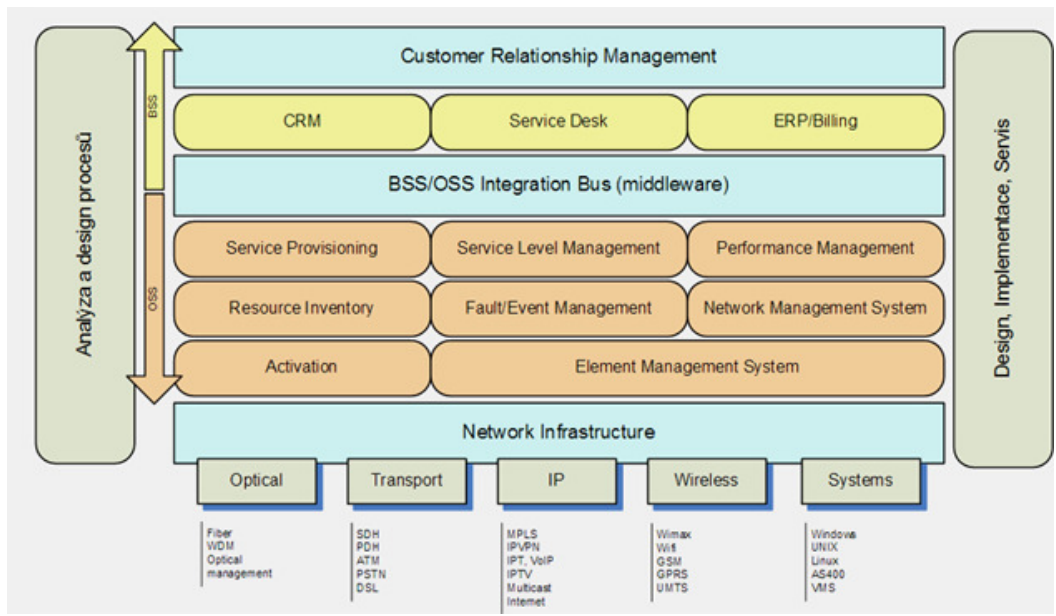

BSS /OSS ensure the proper operation of the telecom organization.

BSS are: CRM, Billing, ERP
OSS are: Ldap, Radius, SSW, MSAN

Concerning the above we have:

- Requirements for Integration between BSS with OSS.
- Requirements for designing and implementing Business Processes For BSS.
- Requirements for Monitoring BSS/OSS.

Orchestration-Routing-Transformation required for Integration and Business Processes.

## 2. SOA

### 2.1.Description & Functionality

**Service-oriented architecture** (**SOA**) is a software architecture design pattern where discrete pieces of software are exposed as services, providing certain functionality to other applications. This is known as Service Orientation. It is vendor, product and technology independent.

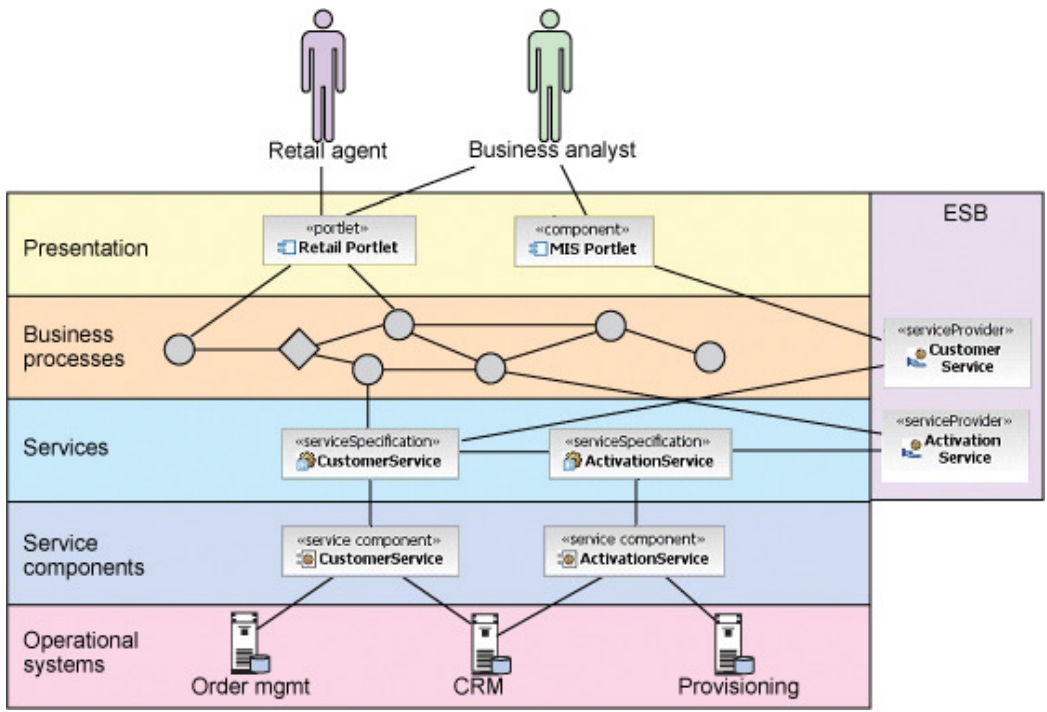

Figure 2.

As shown is Figure 2 a template for a SOA architecture could be:

1. Operational systems layer
    1. Packaged applications
    2. Custom applications
2. Service components layer
    1. Certain Functional areas covered by this layer
3. Services layer
    1. Service components combined and categorized service portfolio is exposed.
4. Business processes layer
    1. Business processes to be represented as choreographies
5. Presentation layer
    1. Web services invoked by Portlets at the user interface level.
6. Integration layer
    1. ESB functionality
    2. Security issues.
    3. Performance issues.
    4. Standards and Technology limitations.
    5. Service management and monitoring.

**Operational systems layer.** Contains either packaged either custom applications as CRM, ERP, business intelligence applications, billing and Service Provisioning. SOA can integrate existing systems with service-oriented techniques integration.

**Service components layer.** Certain Functional areas covered by this layer. These service components handle discrete pieces of functionality under the corresponding SLAs ensuring the necessary Qos of the services.

**Services layer.** Service components combined as composite services. A service portfolio is exposed to be invoked.

**Business process layer.** Invokes the Service Portfolio. A workflow through orchestration rules bundles these services. Each built workflow corresponds to a single application. These workflows support specific business processes.

**Presentation layer.** Web services (WSDL) invoked by Portlets at the user interface level. They cover the functionality described in the Business process layer. These portlets can be accessed by a business analyst or a simple retail agent with the necessary security contstraints (group, roles).

**Integration layer.** Through this layer the integration of services is performed under a set of capabilities, such as mediation, routing, and transformation. All these are covered through the ESB functionality. In this layer decisions for the security and performance issues are taken in order to comply to all necessary standards and agreed SLAs. The service management and the service monitoring is also performed in this layer.

## 2.2 SOA – Conclusions

With SOA
1. We can implement the integration required between BSS /OSS.
2. We can define the business process as described in BSS
3. We can monitor all the activities and operation performed in BSS/OSS

SOA covers the above functionalities through:
1. ESB performs routing, transformation, integration
2. BPM defines business processes
3. EAI (Enterprise Application Integration) performs orchestration, integration (through BPEL)
4. BAM performs activity monitoring

In SOA each layer is built from a separate system and an integration between these systems is required. This means that we need to go into a process of unifying different systems where they are built under different rules. Since these systems can be also as stand alone in the company this means that their union requires subsystems to play the role of adapters between them. These systems have to expose such agnostic and generic interfaces in order to communicate each other thus SOA major rules are loose coupling and reuse. The agnostic and generic interfaces sometimes require tedious and huge implementation. So we realise that integration of these system is not often a simply work to do.

SOA also provides loose coupling across all the services but this also makes more difficult to trace problems.

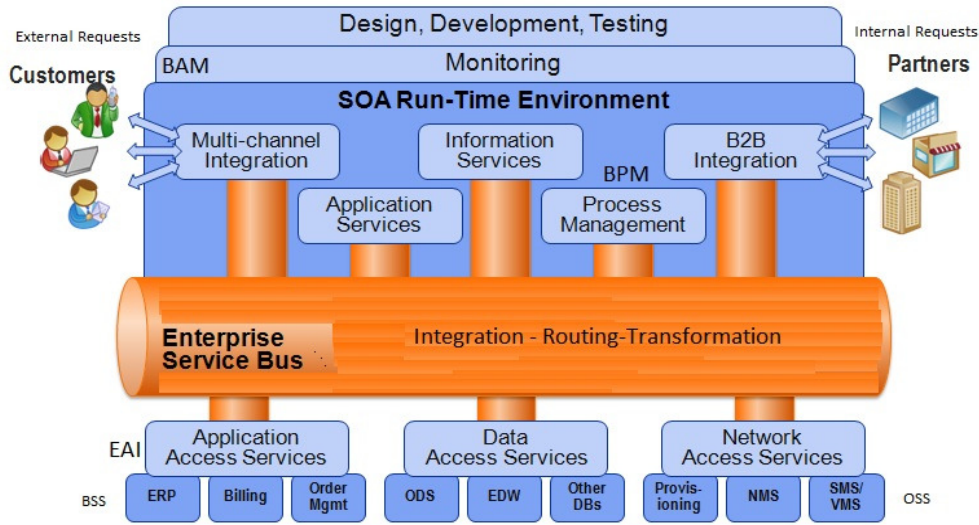

Figure 3.

However, a more closer investigation shows to us, that all these systems have a common feature the "initial request" (we assume that everything is a request) Depending on the system that process the request, this could be an event, service, message or data. So the request is

1. data when the system performs transormation
2. message when the system performs routing
3. event when the system performs monitoring
4. service when the system performs fulfillment

## 3. FRM

Analyzing the Systems and the Layers inside SOA a question is born. Why having all these different systems?

All the above layers of SOA can be implemented with a unified approach.

### 3.1 Description & Functionality

FRM is a software design pattern placed in the layer of fulfillment in the area of BSS/OSS. The main concept is the "request" where depending on the type of claim there is the corresponding fulfillment of this request under an implementation with common rules and unified approach.

With **FRM** we have common aproach to implement ESB, BPM Orhestration, Mediation functionalities. We don't have such a categorization of these. We have only requests and fulfillment of these requests. FRM is one system, with different functions and features according to the requirements referred each time.

With **FRM** we see the running of the business in light of the requests. These requests could be either external (e.g made by a customer to BSS) either Internal (made by sales partner to BSS).

SOA as mentioned by definition is mainly Service Oriented while FRM is Request Oriented. A request can be anything. We can implement SOA using FRM approach.

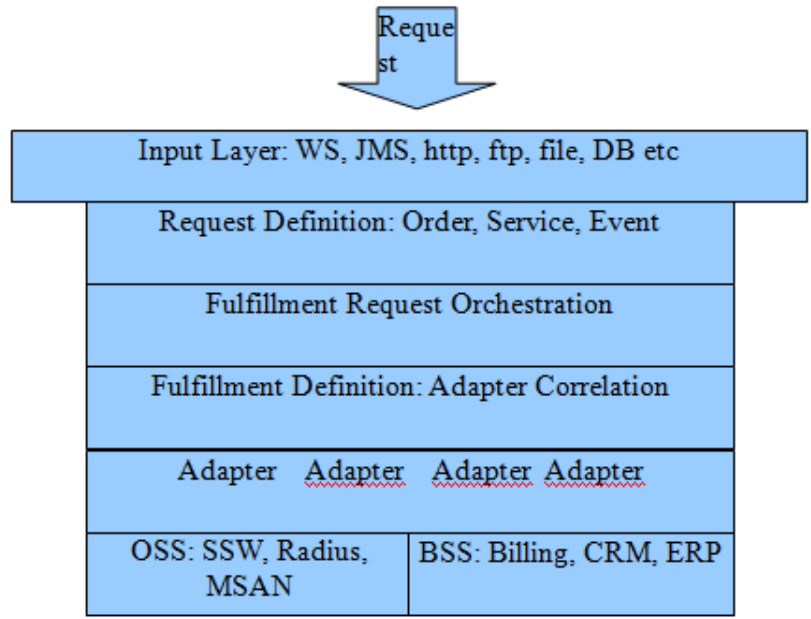

Figure 4

Using a single system (and architecture) FRM advantages and contribution will be:

1. Instead of N different systems we have to choose our choice turns out to be simple and reflects as in one single system. In the market there are many ESB / SOA systems, many order entry, many BPM etc and the optimal combination of these is a tedious, time consuming and costly process. Having one system greatly reduces the search criteria to find the most appropriate solution.
2. Cost Reduction
    a. Usage cost. There is a reduction in the cost of use since instead of 3 or 4 systems we have 1. Therefore the company is not required to buy licenses for all these systems, but only one.
    b. Operation cost. It is easier to manage and operate one system than running N systems. Furthermore it requires fewer people to operate 1 than N systems.
    c. Administration cost. It is easier to administrate one system than N systems. In any case you need fewer people to cover this need.
3. Easy adaptation between systems because talking about one and the same system. The N systems under the umbrella of FRM communicate fully with each other and the connections are intact and unified.
4. Interoperability between processes and services. The FRM is not just one system, but a concept of an approach in which hosted processes, services, events and orders are fully synchronized between them.
5. Reusability. Many of the FRM entities based on reuse depending on usage. The concept of one system alone gives us the possibility of FRM components reused either in processes or in services etc.
6. Easily you can the trace a problem since you have only one system to search for it.

7. Fast Implementation. The concept of a common approach and common architecture creates some standards that help in easy and fast development. For N systems referred, with FRM we have common development philosophy and operation. Consequently this reduces the development time of individual implementations that may be required.
8. Easily export of reports and KPIs. The export reports from the various systems as well as the individual control between the various systems are a painful but necessary process in an enterprise. Through reports we test the systems and produce the necessary data to be used by CRM and then from the management of the company for various purposes: economic, strategic, etc. Through FRM the reports are coming out of one system and there is no complexity in to combine reports from N systems. Reports in FRM are exported through a single common data model which makes reporting easy to use and implement in less time and low probability of error.

### 3.2 Description of FRM model

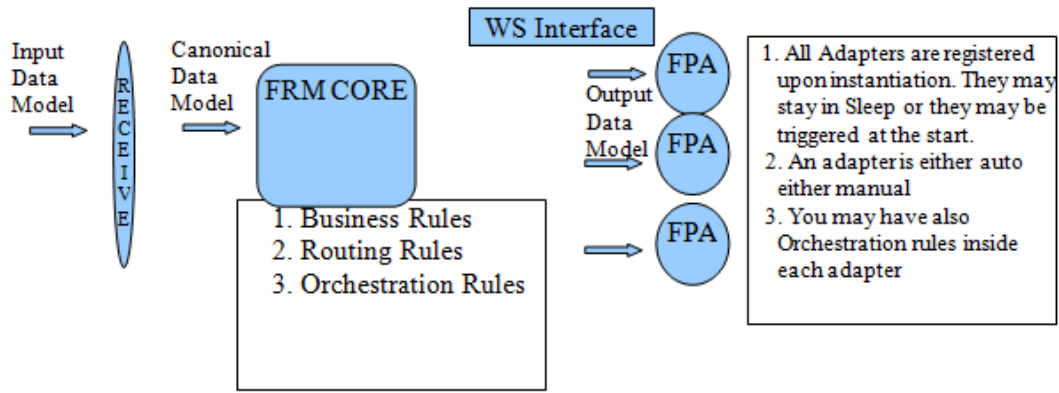

Figure 5

FRM is an architecture which can be divided into three entities as illustrated by the above figure.

1. We have the Receiver who is responsible for entering the data regardless what these are. (order, event, process)
2. FRM Core. The centralized entity where within they are all the rules business, routing and orchestration.
3. Fulfillment Process Adapters which include the third interconnection systems (e.g. it can be an interface with ERP system). Within the FPAs can also define rules orchestration that will affect the actual FPA or another. The reason for this is because we want to be able to extend the orchestration outside the core. They can either in standby mode either pre-activated depending on the rules defined in the orchestration. The FPAs operating simultaneously, synchronously or asynchronously and interact according to the orchestration that will define.

The FRM uses a normalized data model. Step of normalization of data is a necessary step to implement a relational database, as long as results:

a. using all the data needed for system,
b. organizing data in a form that a data exists in one and only one point.
c. visualization of the relationships between the entities of interest to the system.

When entering data in the Receiver they are converted to the normalized model so that all operations and processes within FRM Core are made in a single data model. The normalized model is converted through the Fulfillment Process Adapters in the model that requires the respective third interconnection system.

## 4. CONCLUSIONS

Summing up all the above, the answer to the question why someone should choose a FRM system is:

A FRM system is the interconnection of requirements and procedures within the company. With FRM we see the running of the business in light of the requests. External or Internal depending on the type of claim. As we have mentioned FRM is not simply a system, but a concept, an approach and architecture designed to process claims under common rules, creating a powerful mechanism requirements management within the broader Orchestration processes and requirements.

With FRM we have common aproach to implement ESB, BPM Orhestration, Mediation functionalities. We don't have such a categorization of these.

With FRM a system virtualization is performed and all these systems are implemented through one system under a unified approach.

Instead of N different systems we have to choose, our choice turns out to be simple and reflects as in one single system.

International Journal of Computer Science & Information Technology (IJCSIT) Vol 6, No 1, February 2014

**Authors**

**Stylianos D. Gkoutzioupas**

**Professional Experience:**

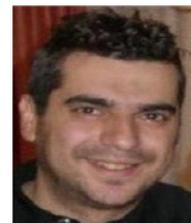

- 2009 – today: CYTA Hellas SA.

 Software Team Leader  (http://www.cyta.gr) : Analysis, design and implementation for Order Management and Service fulfilment applications
- 2008 – 2009: Rayo – e-rhetor

Chief Architect – partner of Rayo (http://www.rayo.gr) : Professional Services for IT applications. Analysis, design and implementation for Provisioning and CRM applications. Working experience on Speech Synethesis(TTS) and Speech Recognition (ASR),
- 2006 – 2008: On Telecoms S.A.

Senior Software Engineer. Designing and developing software IT Services for a 3play operator.
- 2006 – Oct. 2006: INTRACOM Telecommunications Industry S.A

Senior Software Engineer. Designing and developing software applications for Intelligent Networks and IT Services.
- 2006 – 2006 Sept: ATERMON SA (Brussels)

IT Consultant. Analysis and design for Telecom Operators.
- 2001 - 2005:  INTRACOM Telecommunications Industry S.A

Software and Systems Engineer. Designed and developed software applications for postpaid and prepaid billing platforms.
Experience in customer care, product management, team leading.

- 1999 – 2006: NTUA (National and Technical University of Athens)

Participation in several NTUA's European projects. Project analysis and implementation of components.

**Education:**

- 2010 – today: PhD candidate. Research in the area of Fullfiment Management covering SOA, BPM, OM systems (in progress). University of Aegean (School of Business, Department of Business Administration)Michalio building, 8 Michalon Str,, 82100 Chios (Greece)
- 2002 - 2004: National Technical University of Athens (NTUA).

Master in Techno–economic systems.
- 1995 - 2001: National Technical University of Athens (NTUA).

Degree in Electronic Engineering and Computer Science, majoring in Telecommunications Engineering